\documentclass[aip,reprint]{revtex4-1}

\usepackage{graphicx}
\newcommand{\sgn}{\mathop{\mathrm{sgn}}}
\usepackage{physics}
\usepackage{amsfonts}

\begin{document}

\title{Polariton response in the presence of Brownian dissipation from molecular vibrations}

\author{Kalle S. U. Kansanen}
\email[Corresponding author: ]{\url{kalle.s.u.kansanen@jyu.fi}}

\affiliation{Department of Physics and Nanoscience Center, University of Jyv\"askyl\"a, P.O. Box 35 (YFL), FI-40014 University of Jyv\"askyl\"a, Finland}

\author{J. Jussi Toppari}
\affiliation{Department of Physics and Nanoscience Center, University of Jyv\"askyl\"a, P.O. Box 35 (YFL), FI-40014 University of Jyv\"askyl\"a, Finland}

\author{Tero T.~Heikkil\"a}
\affiliation{Department of Physics and Nanoscience Center, University of Jyv\"askyl\"a, P.O. Box 35 (YFL), FI-40014 University of Jyv\"askyl\"a, Finland}

\date{\today}

\begin{abstract}
We study the elastic response of a stationarily driven system of a cavity field strongly coupled with molecular excitons, taking into account the main dissipation channels due to the finite cavity linewidth and molecular vibrations. We show that the frequently used coupled oscillator model fails in describing this response especially due to the non-Lorentzian dissipation of the molecules to their vibrations. Signatures of this failure are the temperature dependent minimum point of the polariton peak splitting, uneven polariton peak height at the minimum splitting, and the asymmetric shape of the polariton peaks even at the experimentally accessed "zero-detuning" point. Using a rather generic yet representative model of molecular vibrations, we predict the polariton response in various conditions, depending on the temperature, molecular Stokes shift and vibration frequencies, and the size of the Rabi splitting. Our results can be used as a sanity check of the experiments trying to "prove" results originating from strong coupling, such as vacuum-enhanced chemical reaction rate.
\end{abstract}


\maketitle

\section{Introduction}

Strong coupling between electromagnetic modes and electronic excitations has emerged as a tool to modify internal material properties and dynamics in various systems, from light harvesting~\cite{groenhof2018coherent} and energy transport~\cite{coles2014polariton,zhong2016non,zhong2017energy,akulov2018long} to controlling photochemical reactions~\cite{hutchison2012modifying,munkhbat2018suppression,ribeiro2018polariton, feist2017polaritonic}. The strong coupling regime is typically reached when the  interaction-driven splitting of the dressed state, or polariton, eigenenergies becomes larger than their linewidth. 
Only then, the avoided crossing in the energy spectrum can be identified. This is rather clear in systems with Lorentzian response that can be described by adding imaginary parts to the photon/exciton eigenenergies. However, in the case of molecules, the main cause of linewidth broadening often originates from vibrational dissipation, which does not typically produce a Lorentzian response. Here we consider in detail a rather generic model system (two-level system and harmonic vibrations) to explore the effects of Brownian vibrational dissipation to the polaritonic spectrum. Comparing the results of such a model to the experimentally obtained polariton fingerprints then provides a sanity check of those experiments, helping to rule out spurious effects.

Our approach is based on the open quantum system model that we introduced in Ref.~\onlinecite{kansanen19} to study the polariton response in the case where surface plasmon polaritons couple strongly with molecular excitations. Here we modify this approach to concentrate on the case of a cavity containing a large number of molecules interacting with the cavity mode. The qualitative difference between these systems is that in the cavity case the molecular excitations are coupled with the free field only via the cavity field. We also focus on studying the role of vibrational dissipation on renormalizing the effective strong coupling parameters, especially in the case approaching the overdamped vibrations. 

\begin{figure}
    \centering
    \includegraphics{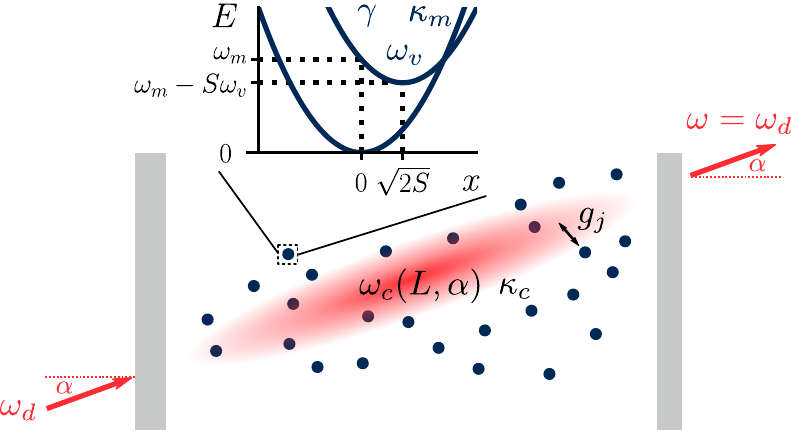}
    \caption{Schematic picture of a measurement setup in which the cavity eigenfrequency $\omega_c$ is controlled by both the length $L$ between the mirrors and the angle $\alpha$ of incident light of frequency $\omega_d$. This confined light mode couples to molecules with harmonic vibrations. In this article, we focus on the elastic cavity emission or transmission, that is, the observed light power at frequency $\omega = \omega_d$. For explanation of other parameters, see the main text.}
    \label{fig:setup}
\end{figure}

We consider the setup shown in Fig.~\ref{fig:setup}. It consists of a cavity with eigenfrequency $\omega_c$ and linewidth $\kappa_c$, probed externally via a field coupling through one of the cavity mirrors. Inside the cavity, there are a (large) number $N$ of molecules with excitation frequency $\omega_m$. 
We denote the coupling energy between the cavity fundamental mode and a single exciton by $g_j$.
At the same time, the molecular excitation couples to its vibration mode with eigenfrequency $\omega_v$ via a coupling strength $\sqrt{S}\omega_v$, where $S$ is the Huang-Rhys factor~\cite{huang1950theory}. We assume the vibrations to reside in a bath with temperature $T$, providing them with a linewidth $\gamma$. This vibrational coupling gives the individual molecules their Stokes shift and provides their inhomogeneous broadening~\cite{mahan2013many}. 

\begin{figure}
    \centering
    \includegraphics{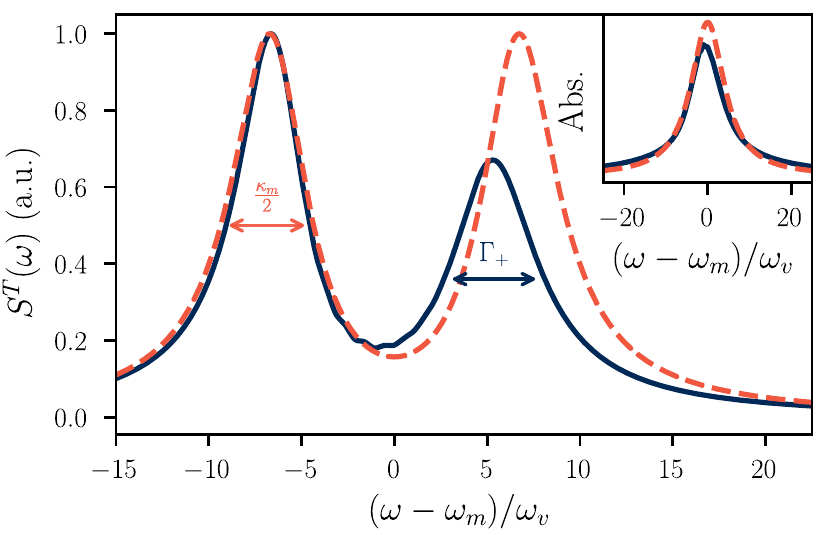}
    \caption{Example of the polariton spectra under the coupled oscillator model (orange dashed line) and our $P(E)$ model (blue solid line), assuming a very high finesse cavity tuned so that the spacing between the polariton peaks is at its smallest. Inset shows the respective absorption profiles. Since the cavity dissipation $\kappa_c$ is negligible, in the Lorentzian model we find Lorentzian polariton peaks whose linewidth is half of the ''molecular'' linewidth $\kappa_m$. When the vibrations and their dissipation are taken into account, one can find non-Lorentzian behaviour, renormalization of Rabi splitting, as well as changes in the linewidths. }
    \label{fig:comp}
\end{figure}

We illustrate the effects of this inhomogeneous broadening on the polariton eigenmodes via the system response in Fig.~\ref{fig:comp}. 
Compared to the model in which the molecule response is taken as Lorentzian --- known as the coupled oscillator model --- the polaritonic spectrum has a few distinct features.
First, the vibrations cause an asymmetry between the upper and lower polariton peak which is similar to that obtained by changing the detuning between the cavity and molecular frequency.
However, this effect cannot be imitated by using a Lorentzian molecular response because this asymmetry is caused by the asymmetry of the molecular response.
Second, we find that the polaritonic frequencies are renormalized, which affects the observed Rabi splitting.

\subsection{Coupled oscillators}
We briefly motivate the upcoming discussion with the often-used ''coupled oscillators'' model of polaritonics.
It describes the cavity and excitonic modes as effectively harmonic oscillators; cavity mode has the eigenfrequency $\omega_c$ while for the exciton it is $\omega_m$.
The rotating wave approximation is often made to simplify the coupling between these modes~\cite{walls2007quantum}.
Then, in the single-excitation subspace the Hamiltonian for a single molecule in a cavity is given by ($\hbar = 1$ throughout the text)
\begin{align}
    H = \mqty(\omega_c & g \\ g^* & \omega_m).  
\end{align}
This is the Jaynes-Cummings Hamiltonian~\cite{jaynes1963comparison}.
The polariton eigenfrequencies are then obtained by diagonalizing this matrix
which gives
\begin{align}
    \omega_\pm = \frac{\omega_c + \omega_m }{2}\pm \sqrt{\abs{g}^2 + (\omega_c - \omega_m)^2/4}.
    \label{eq:intro:polfreq1}
\end{align}
Perhaps the most straightforward and often-used way to model dissipation is to introduce an imaginary shift in the eigenfrequencies.
That is, we set $\omega_c \rightarrow \omega_c - i \kappa_c$ for the cavity and similarly $\omega_m \rightarrow \omega_m - i \kappa_m$ for the exciton.
This method leads to Lorentzian lineshapes for the cavity and the exciton alone.
Consequently, if the dissipation is inserted to the polaritonic frequencies at resonance $\omega_c = \omega_m$, one finds
\begin{align}
    \omega_\pm = \omega_m - i \frac{\kappa_c + \kappa_m}{2} \pm \sqrt{\abs{g}^2 - (\kappa_c - \kappa_m)^2/4},
    \label{eq:intro:polfreq2}
\end{align}
from which one infers $(\kappa_c + \kappa_m)/2$ to be the polariton linewidth~\cite{torma2014strong}.

However, the molecular vibrations cannot often be neglected in molecular polaritonics.
It is quite obvious that the vibrational modes have their own complex dynamics as well as coupling to their environment which shape the absorption and fluorescence spectra of molecules.
We show that even in the simplest vibrational models the dissipative properties of the molecular vibrations, which in fact are the main source of line broadening, have quite intricate physics and thus effects on the polariton spectrum.

\section{Input-output theory of molecule-cavity spectroscopy}

In this article, we consider a simplified model of a molecule as a two level system with harmonic vibrations.
In the language of operators, we associate a lowering operator $\sigma$ to the electronic two level system while the phonons of vibrations are destroyed by $b$.
The Hamiltonian is then characterized by the eigenenergies $\omega_m$ and $\omega_v$ of the two level system and vibrations, respectively, and the dimensionless Huang-Rhys factor $S$ quantifying their coupling
\begin{align}
    H_\mathrm{mol} = \omega_m \sigma^\dagger \sigma + \omega_v b^\dagger b + \omega_v\sqrt{S} \sigma^\dagger \sigma (b + b^\dagger).
\end{align}
We assume that there are $N$ identical particles which are described by this Hamiltonian.

Here, we consider that these molecules are embedded in a Fabry-Perot cavity of eigenfrequency $\omega_c$. 
For instance, the cavity eigenfrequency may be tuned by controlling the length of the cavity $L$ and the incident light angle $\alpha$ by the relation $\omega_c = \frac{\pi c}{L}/\sqrt{1 - \sin^2(\alpha)/n_\mathrm{eff}^2}$ where $c$ is the speed of light and $n_\mathrm{eff}$ the effective refractive index inside the cavity~\cite{skolnick1998strong}.
The electronic coupling strength is described by a constant $g_j$.
When we denote the cavity photon annihilation operator by $c$, the full Hamiltonian under the rotating wave approximation is
\begin{align}
    H = \omega_c c^\dagger c + \sum_{j=1}^N \qty(H_{\mathrm{mol},j} + g_j c^\dagger \sigma_j + g_j^* \sigma_j^\dagger c),
\end{align}
which is often referred as the Holstein-Tavis-Cummings Hamiltonian~\cite{herrera2016cavity}.
Under the assumption of identical particles, the relevant strong coupling constant is $\sum_{j=1}^N \abs{g_j}^2 \equiv g_N^2$.

In order to observe anything spectroscopically, the cavity must be driven by an external light source which we assume to have a low power so that we can concentrate on the lowest order response.
We assume a laser drive at a driving frequency $\omega_d$.
Then, some light will leak out of the other side of the cavity.
We are interested in this transmission spectrum $S^T(\omega; \omega_d)$ where, in general, $\omega$ is the frequency of the transmitted light.
Any observed polaritonic emission is mediated by the cavity --- there is no direct emission from the molecules.
This simplifies the description of the spectrum as there cannot be interference between the cavity and molecular emission.
Also, the light emission should mostly be observed at the driving frequency.
In this article, we neglect all processes that could cause inelastic behaviour.
This may be experimentally guaranteed by fixing the outgoing angle to be the same as the incident angle and measuring light power only at the frequency equal to the driving frequency $\omega = \omega_d$.

We assume that there exists an environment for each of the vibrational, excitonic, and cavity modes; coupling to these environments leads to dissipation.
Both the driving and dissipation may be taken into account in the input-output formalism of quantum optics~\cite{gardiner1985input}.
It describes the quantum dynamics in the Heisenberg picture and can be considered an open quantum system modification to the Heisenberg equation.
Without vibrations, this method gives similar results to those alluded in Eq.~\eqref{eq:intro:polfreq2}.
However, the vibrations complicate finding the cavity reflection and transmission spectra notably
as the equations of motion are nonlinear.
The solution is obtained by moving into polaron frame, i.e., finding the dynamics of $\sigma^S = \sigma e^{\sqrt{S}(b^\dagger - b)}$, 
which allows for simplifying approximations that in the end decouple the vibrational dynamics from those of the cavity-exciton system~\cite{kansanen19,reitz2019langevin}.
The cavity transmission or emission spectrum is in this model
$S^T(\omega; \omega_d) \propto \abs{r(\omega_d)}^2\delta(\omega - \omega_d)$
where the physics is contained in the cavity response function~\cite{kansanen19}
\begin{equation}
	r(\omega_d) = \qty[i(\omega_d - \omega_c) - \frac{\kappa_c}{2} +  g_N^2 A(\omega_d - \omega_m + S\omega_v)]^{-1}.
    \label{eq:response}
\end{equation}
The molecular contribution to the polaritonic spectrum is the absorption function 
\begin{equation}
	A(\omega) = \int \dd{E} \frac{P(E)}{i(\omega - E) - \kappa_m/2}.
    \label{eq:A}
\end{equation}
The absorption profile of the molecule without the cavity is given by $\Re[-A(\omega)]$.
Here, $P(E)$ describes the probability to emit ($E>0$) or absorb ($E<0$) energy $E$ to/from the vibrations.
It can be expressed as
\begin{equation}
    P(E) = \int \frac{\dd{t}}{2\pi} e^{i E t} \expval{e^{\varphi(t)}e^{-\varphi(0)}}
    = \int \frac{\dd{t}}{2\pi}  e^{i E t} e^{J(t)-J(0)},
    \label{eq:generalPE}
\end{equation}
where $\varphi(t)=\sqrt{S} [b^\dagger(t)-b(t)]$ is proportional to the momentum operator of the vibrations in the Heisenberg picture, while $J(t) = \expval{\varphi(t)\varphi(0)}$ is proportional to momentum correlator. 
The Fourier convention has been chosen this way to allow for a probability interpretation: the energy integral over $P(E)$ amounts to unity.
Furthermore, the same formalism extends to many vibrational modes: the total $P(E)$ in Eq.~\eqref{eq:A} is then a convolution of all the single-mode $P(E)$ functions~\cite{kansanen19}.

The expression of $A$ is the convolution of the vibrational $P(E)$ and the electronic susceptibility.
The latter alone would produce a Lorentzian lineshape with linewidth $\kappa_m$.
This linewidth 
follows from the assumption of excitonic dissipative environment which does not couple to the observed far-field mode. 
It is renormalized by the coupling to vibrations as $\kappa_m = \tilde{\kappa} + S \gamma$ where $\tilde{\kappa}$ represents the dissipation rate of the electronic transition and $\gamma$ is the dissipation rate of the vibrations~\cite{kansanen19}.
However, $P(E)$ ultimately determines the lineshape in the presence of vibrations.

If there are no vibrations, $S = 0$ and $P(E) = \delta(E)$, the solution of $1/r(\omega) = 0$ gives exactly the polariton frequencies in Eq.~\eqref{eq:intro:polfreq2}.
However, in this context, the frequencies are real.
When $\omega_c, \omega_m, \abs{g} \gg \kappa_c, \kappa_m$, one can first neglect the dissipation rates and minimize the imaginary part to find the polariton frequencies in Eq.~\eqref{eq:intro:polfreq1}.
It is then straightforward to show that the behaviour around these eigenfrequencies is Lorentzian and the real part gives the dissipation rate $(\kappa_c +  \kappa_m)/2$.
The proper way to find the polariton peaks would be to find the extremal points of $\abs{r(\omega)}^2$ which gives the dissipative correction. 
A closed form solution is possible to obtain only at resonance $\omega_c = \omega_m$. 
It reads
\begin{equation}
    \omega_\pm = \omega_m \pm \sqrt{ \abs{g}^2 \sqrt{1 + \frac{\kappa_m(\kappa_m+\kappa_c)}{2\abs{g}^2}} - \kappa_m^2/4}.
\end{equation}
The square root term may be approximated by $\sqrt{\abs{g}^2 + \kappa_c \kappa_m/4}$ when $\abs{g} \gg \kappa_m, \kappa_c$.
This result differs from the coupled oscillator model in two ways: First, dissipation strictly increases the spacing between polariton frequencies, or the Rabi splitting, when the coupling is large compared to the dissipation rates.
This changes at the opposite limit when the dissipation rates dominate the coupling $\abs{g}^2$, leading eventually to the disappearance of the polariton peaks.
Second, the dissipation in the cavity and the molecule are, in general, not interchangeable as in the coupled oscillator model.
This is a direct consequence of our assumptions: we only couple the external light source to the cavity and observe only its emission.

\section{$P(E)$ theory under Brownian dissipation}
\label{sec:PofE}

We now derive the $P(E)$ function assuming that the molecular vibrations are coupled to a bath of harmonic oscillators.
This means solving the correlator $\expval{e^{\varphi(t)}e^{-\varphi(0)}}$ and then evaluating its Fourier transform.
We note the recent similar approaches of Refs.~\onlinecite{reitz2020molecule,clear2020phonon} that model molecules embedded in a crystal taking into account coupling between the lattice vibrations and molecular vibrations.

The correlator $\expval{e^{\varphi(t)}e^{-\varphi(0)}}$ can be evaluated for different models~\cite{kansanen19,reitz2020molecule,reitz2019langevin} so let us make a general argument:
since the vibrations are harmonic, their fluctuations follow Gaussian statistics.
Furthermore, we assume that these fluctuations are stationary, i.e., all correlators depend only on time differences and not on any specific time.
By introducing a time ordering operator $\mathcal{T}$ that always orders $\varphi(t)$ before $\varphi(0)$ we can write the correlator as $\mathcal{T}\expval{e^{\varphi(t)-\varphi(0)}}$.
This may be identified as the characteristic function of the stochastic quantity $\varphi(t)-\varphi(0)$, which has a vanishing mean.~\cite{terosbook,ingoldnazarov}.
In the case of Gaussian fluctuations, the correlator is fully determined by the variance $\mathcal{T}\expval{[\varphi(t)-\varphi(0)]^2}/2$ which equals to $J(t) - J(0)$.

Because vibrational modes are often low frequency compared to optical frequency, the quantum optical models of dissipation are not justified.
This is because in the quantum optical case the time scales of the system are generally assumed to be much smaller than its relaxation time.
However, this is not a typical limit for molecular vibrations.
This leads to a failure of the rotating wave approximation that is made to the system-environment coupling in the quantum optical formulation~\cite{breuer2002theory}.
In the Brownian case (also called the Caldeira-Leggett model~\cite{caldeira1984influence}), the equations of motion for the position $x$ and momentum $p$ are
\begin{align}
    \dot x = \omega_v p, \quad \dot p = -\omega_v (x + \sqrt{2 S}\sigma^\dagger \sigma) - \gamma p + \xi,
    \label{eq:clm:eom}
\end{align}
where $\xi$ is a Langevin force of thermal fluctuations obeying a noise correlator~\cite{giovannetti2001phase}
\begin{align}
    \expval{\xi(\omega)\xi(\omega')} = \frac{\gamma \omega}{2\pi \omega_v}\qty[\coth(\frac{\omega}{2k_B T}) + 1]\delta(\omega + \omega').
    \label{eq:noisecorr}
\end{align}
On average the molecules are in their ground state, since $\omega_m \gg k_B T$, so we neglect the $\sigma^\dagger \sigma$ -term in Eq.~\eqref{eq:clm:eom}.
This assumption decouples vibrational dynamics from those of the exciton.
The solution of $x$ and $p$ can then be obtained in the Fourier space in terms of the force $\xi$.

By using $\varphi(t) = \sqrt{2 S} p(t)$ and taking a few algebraic steps we find an integral expression
\begin{equation}
    \begin{split}
    J(t)  &=\frac{S \gamma}{\pi \omega_v} \int \dd{\omega} e^{-i\omega t}
    \frac{\omega^3}{(\omega^2-\omega_v^2)^2+\omega^2 \gamma^2} \\
    &\phantom{\frac{S \gamma}{\pi \omega_v} \int \dd{\omega} e^{-i\omega t}}
    \times\qty[\coth(\frac{\omega}{2k_B T}) + 1 ]. \label{eq:Jintegral}
    \end{split}
\end{equation}
It should be noted that this integral does not converge when $t = 0$, that is, the variance of momentum diverges.
This is akin to the free Brownian particle for which the variance of the position diverges.
We may solve the integral for $t \neq 0$ and deal with this divergence at a later point.

Whereas Ref.~\onlinecite{kansanen19} computes the $P(E)$ in the limit $\gamma \ll \omega_v$, here we solve the problem with an arbitrary $\gamma/\omega_v$.
We employ the method of residues to solve the integral for $t > 0$ by choosing an infinite radius semi-circle in the lower complex half-plane as the integration contour.
Formally, we then set $J(t) - J(0) \equiv \tilde{J}(t)$ to vanish at time $t = 0$ and cut off the divergence.
Lastly, we can use the stationarity of vibrations to expand to negative times by the relation $J(-t) = J(t)^*.$
These steps are taken to ensure the consistency of $P(E)$.

\begin{figure}
    \centering
    \includegraphics{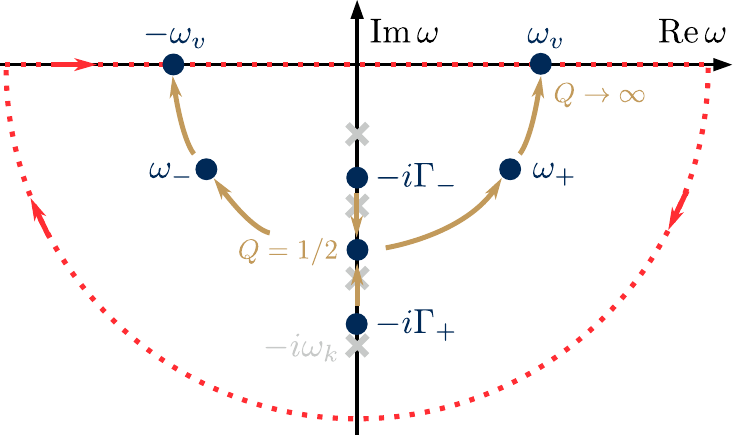}
    \caption{Sketch of the singularities of $g(\omega)$ (blue circles) in the lower complex half-plane as a function of the quality factor $Q$. 
    The singularities of $\coth(\omega/2k_B T)$ that lie on the imaginary axis are depicted by gray crosses.
    The red dotted line represents the integration contour for $t>0$.
    This residue structure is mirrored in the upper complex half-plane.}
    \label{fig:intcontour}
\end{figure}

The integral in Eq.~\eqref{eq:Jintegral} may be separated into two contributions by the singularities of the integrand that lie inside the integration contour (see Fig.~\ref{fig:intcontour}).
On one hand,  there is the contribution of the singularities of the rational function $g(\omega) \equiv  \omega^3/[(\omega^2-\omega_v^2)^2+\omega^2 \gamma^2]$ which is determined by the quality factor $Q=\omega_v/\gamma$ of vibrations.
Especially, when $Q > 1/2$ there are two singularities with non-zero real frequencies which are associated with underdamped motion.
The value $Q = 1/2$ represents the critical damping of harmonic motion, while $Q < 1/2$ corresponds to overdamped motion with two singularities on the imaginary axis.
On the other hand, the hyperbolic cotangent has an infinite series of singularities at complex frequencies which are related to the temperature. 
This we term \emph{the Matsubara contribution}.

Since we want to calculate the Fourier transform of $P(t) = \exp[\tilde{J}(t)]$, it is useful to use the convolution theorem.
It allows us to Fourier transform the individual components (each corresponding to one singularity/residue) and convolving the Fourier transforms together only in the end.

The calculation is tractable due to a group property of Lorentzian distributions under convolutions.
We find that the Fourier transforms may be written in terms of two functions
\begin{subequations}
\begin{align}
    f_L(\omega; \omega_0, \Gamma) = \frac{1}{\pi} \frac{\Gamma}{(\omega-\omega_0)^2 + \Gamma^2}, \\
    g_L(\omega; \omega_0, \Gamma) = \frac{1}{\pi} \frac{\omega - \omega_0}{(\omega-\omega_0)^2 + \Gamma^2}.
\end{align}
\end{subequations}
We recognize $f_L$ to be the Cauchy-Lorentz probability distribution function and $g_L$ to be its Hilbert transform.
We use these normalized functions as they integrate to unity ($f_L$) or to zero ($g_L$) for all parameters.
Also, these functions follow the convolution table 
\begin{subequations} \label{eq:convrules}
\begin{align}
    [f_L(\omega_1,\Gamma_1) \ast f_L(\omega_2,\Gamma_2)](\omega) = f_L(\omega; \omega_1 + \omega_2, \Gamma_1 + \Gamma_2) \\
    [g_L(\omega_1,\Gamma_1) \ast f_L(\omega_2,\Gamma_2)](\omega) = g_L(\omega; \omega_1 + \omega_2, \Gamma_1 + \Gamma_2) \\
    [g_L(\omega_1,\Gamma_1) \ast g_L(\omega_2,\Gamma_2)](\omega) = -f_L(\omega; \omega_1 + \omega_2, \Gamma_1 + \Gamma_2).
\end{align}
\end{subequations}
Often, only the Lorentzian component is taken into account.
However, the role of $g_L$ is to provide the emission-absorption asymmetry to $P(E)$. 
This general physical rule simply states that temperature (of the environment) dictates the ratio of probabilities between absorption and emission of energy $E$ between the environment and the vibrations.
To be more exact, this ratio is given by $P(E)/P(-E) = e^{-\beta E}$.
This relation follows from the definition of $P(E)$ as a Fourier transform of $\expval{e^{\varphi(t)}e^{-\varphi(0)}}$ by using the definition of Heisenberg operators, cyclic property of trace, and that the vibration mode is in thermal equilibrium~\cite{ingoldnazarov}.

Next, we find the contributions of individual singularities to $J(t)$ and then calculate their Fourier transforms.
The algebraic details of the Fourier transform are given in Appendix~\ref{app:FT}.

\subsection{Residues of the rational function $g(\omega)$} \label{sec:gw}
Let us consider underdamped motion with $Q > 1/2$.
We then find singular points at $\omega = \pm \tilde{\omega}_v \pm i \frac{\gamma}{2}$.
The frequency $\tilde{\omega}_v = \omega_v \sqrt{1 - \frac{1}{4Q^2}}$ is the renormalized vibrational frequency.
By the method described above, we find for all times~$t$
\begin{align}
    \tilde{J}_\pm(t) = \qty[\Re(D_\pm) + i \Im(D_\pm) \sgn{t}] \qty(e^{\mp i \tilde{\omega}_v t -\frac{\gamma}{2}\abs{t}} - 1),
\end{align}
where $\sgn{t}$ is the sign function ($\sgn{t} = t/\abs{t}$ for $t \neq 0$ and $\sgn{t} = 0$ if $t = 0$) and
\begin{subequations}
\begin{align}
    D_- &= S\frac{\frac{i}{Q} - \frac{\omega_v}{\tilde{\omega}_v} \qty(\frac{1}{2Q^2} - 1)}{e^{\beta\qty(\tilde{\omega}_v + i \frac{\gamma}{2})} - 1} 
    \equiv SN \qty[\frac{i}{Q} - \frac{\omega_v}{\tilde{\omega}_v} \qty(\frac{1}{2Q^2} - 1)],\\
    D_+ &= S(N^* + 1)\qty[-\frac{i}{Q} - \frac{\omega_v}{\tilde{\omega}_v} \qty(\frac{1}{2Q^2} - 1)].
\end{align}
\end{subequations}
The constant $N$ defined in $D_-$ seems to be a complexification of the Bose function. 
If we set $\gamma = 0$, $N$ would be the number $n_\mathrm{th}$ of thermal excitations.
One can then readily associate $D_+$ to emission and $D_-$ to absorption.

The Fourier transform of $\exp[\tilde{J}_\pm(t)]$ is given by the following series representation
\begin{widetext}
\begin{align}
    P_\pm(E) = \mathcal{F}\qty[\exp[\tilde{J}_\pm(t)]](E)
    = e^{- \Re(D_\pm)}\sum_{n = 0}^\infty \frac{1}{n!} \abs{D_\pm}^n 
    \qty[\cos(\phi^\pm_n) f_L\qty(E;\pm n \tilde{\omega}_v, n \frac{\gamma}{2}) + \sin(\phi^\pm_n)g_L\qty(E;\pm n \tilde{\omega}_v, n \frac{\gamma}{2}) ],
    \label{eq:Pplusminus}
\end{align}
\end{widetext}
where $\phi^\pm_n = \Im(D_\pm) - n \arg(D_\pm)$ is an angle that depends directly on the complex phase of the coefficient $D_\pm$.
This series represents all the possible vibronic peaks: $P_+$ describes all the processes where a certain number of vibronic excitations are created when the molecule is excited while $P_-$ describes anti-Stokes processes of exciting the molecule from an excited vibrational state.
Since we assume harmonic vibrations, there is no limit to the number of vibrational states but accessing higher vibrational states is unlikely because of the dissipation.
Note that without the angle dependence, $\phi^\pm_n = 0$, which happens when $\gamma = 0$ or $Q \rightarrow \infty$, the prefactors would follow the Poisson distribution with $D_+ = S( n_\mathrm{th} +1)$ and $D_- = S n_\mathrm{th}$.
Thus, the non-zero $\phi^\pm_n$ can be regarded as a sign of correlation between processes that involve multiple quanta of vibration.

In the overdamped limit, $Q < 1/2$, only a few changes are required in Eq.~\eqref{eq:Pplusminus}.
First, there is no harmonic motion and thus $\mp \tilde{\omega}_v \rightarrow 0$.
We rather have two dissipation rates so that 
\begin{align}
    \frac{\gamma}{2} \rightarrow \Gamma_\pm = \omega_v \sqrt{\frac{1}{2Q^2} - 1 \pm \frac{1}{Q}\sqrt{\frac{1}{4 Q^2} - 1}}.
\end{align}
Also, the coefficients change to
\begin{align}
    D_\pm \rightarrow \frac{S}{2Q}\qty[- i + \cot(\frac{ \Gamma_\pm}{2 k_B T})]\qty(1 \pm Q\sqrt{\frac{1 - 2Q^2}{1 + 2 Q^2}})
\end{align}
assuming that the singularities of $g(\omega)$ do not occur at the same point as those of the hyperbolic cotangent. Such coincidences would take place only at numerable points of continuous variables. We hence expect such double poles to have no observable consequences.

\subsection{Matsubara contribution} 
\label{sec:matsubara}
The function $\coth(\omega/(2 k_B T))$ has singularities at $\omega = i2 \pi k_B T k$ for $k \in \qty{1,2,\dots}$.
For any $k$, the contribution to $\tilde{J}(t)$ is given by
\begin{align}
    \tilde{J}_k(t) = C_k\qty(e^{- \omega_k \abs{t}} - 1), C_k = \frac{4 S}{\beta Q}\frac{\omega_k^3}{(\omega_k^2 + \omega_v^2)^2 - \gamma^2\omega_k^2}.
\end{align}
Since there is no imaginary part in $\tilde{J}_k(t)$, its Fourier transform is simply a Lorentzian.
The resulting $P_k$ is then a series of $E = 0$ centered Lorentzians with a width that is multiples of the Matsubara frequency $\omega_k$.
Furthermore, it admits to a representation in terms of the incomplete Gamma function\footnote{Defined as $\Gamma(z;a,b) = \int_a^b \dd{u} u^{z-1} e^{-u}$.} as
\begin{align}
    &P_k(E) = \mathcal{F}\qty[\exp[\tilde{J}_k(t)]](E) \\
    &= e^{-C_k}\qty(\delta(E) - \Re\qty[\frac{(-C_k)^{i \frac{E}{\omega_k}}}{\pi \omega_k}\Gamma\qty(- i \frac{E}{\omega_k}; -C_k, 0)]). \notag
    \label{eq:Pmatsubara}
\end{align}
The Matsubara contribution only widens the $P(E)$.

When we consider the sum $\sum_k \tilde{J}_k(t)$, we encounter a logarithmic divergence as the sum behaves asymptotically as the harmonic series $\sum_k 1/k$.
This is a well known feature of Ohmic dissipation~\cite{leggett1987dynamics} which is characterized by the asymptotically linear $\omega$ dependence of the noise correlator in Eq.~\eqref{eq:noisecorr}.
Thus, we must introduce a cutoff to the integral~\eqref{eq:Jintegral} which we may consider at the level of residues.
Since the divergence is logarithmic, the results depend only weakly on the chosen cutoff.
In this context, the Matsubara frequency $\omega_k$ is also the width parameter of $P_k$ so that the corresponding distribution becomes wider and wider as higher and higher Matsubara frequencies are considered.
Here, we choose the number of Matsubara modes by first choosing some cutoff frequency $\omega_{L}$ and then calculating $k_\mathrm{max} = \left \lfloor{\omega_L/(2\pi k_B T)}\right \rfloor$.
For very low temperatures corresponding to $\omega_v/k_B T \gg 2\pi$ one may also choose to approximate the $k$-sum as an integral.
However, this is not the limit we consider below.
We set $\omega_v/k_B T \in [0.5; 4]$ and choose the cutoff frequency to be $\omega_L = 25 \omega_v$ in our numerical analysis.

\subsection{Absorption function}
Using the results of Sec.~\ref{sec:gw} and~\ref{sec:matsubara}, we may now express the absorption function as a convolution 
\begin{align}
    A(\omega) = [P_+ \ast P_- \ast P_{1} \ast \dots \ast P_{k_\mathrm{max}} \ast \chi] (\omega)
\end{align}
with the electronic susceptibility $\chi(\omega) =  [i \omega - \kappa_m/2]^{-1}$ defined in Eq.~\eqref{eq:A}.
With the value of $A$ it is then straightforward to calculate the elastic spectrum $S^T(\omega)$ using the response function $r(\omega)$ of Eq.~\eqref{eq:response}.

Due to associativity of convolution, we may change the order of convolutions in $A$.
This is numerically useful, as $\chi(\omega) = -\pi[f_L(\omega;0,\kappa_m/2) + i g_L(\omega;0,\kappa_m/2)]$, the susceptibility $\chi$ provides a constant to all the width parameters.
By dividing the susceptibility into parts using the convolution rules, especially all the delta functions in Eq.~\eqref{eq:Pplusminus} acquire a finite width proportional to $\kappa_m$.
This facilitates a numerical method for the calculation of the convolutions, although it is possible also analytically.
There are only a few numerical issues to be aware of: one must have a dense enough discretization of frequencies so that the peaks are well represented and a large range of values must be included so that there are no spurious edge effects in the calculation of convolutions~\cite{numericalcode, harris2020array, virtanen2020scipy}.

Without numerical analysis, one can already have some insight to $P(E)$ and the resulting absorption function $A$.
The effect of the Huang-Rhys factor $S$ is two-fold: because the prefactors in $P_\pm$ are similar to those of Poisson distribution, increasing $S$ increases the support of the $P(E)$ function, i.e., there are more terms in the sum that differ appreciably from zero.
Also, increasing $S$ increases the effective dissipation rate $\kappa_m$ in the molecular susceptibility $\chi$ which further widens the absorption function $A$ and increases the linewidth of individual vibronic peaks.
The role of temperature $T$ is similar to this because it allows for thermal excitations.
Then, the molecular linewidth increases as temperature is increased since there are more processes available which result in the same amount of energy absorbed.
The role of the Matsubara contribution is essentially to be a correction to this thermal broadening.
In conjunction with $S$ and $T$ the quality factor of $Q$ then determines whether individual vibronic peaks may be observed and what the molecular linewidth is.

The derivation of $P(E)$ and consequently of $A$ is here done for a single vibrational mode.
As mentioned below Eq.~\eqref{eq:generalPE}, the definition of $P(E)$ extends to many vibrational modes through convolution.
One can then identify each element of $P(E)$ ($P_\pm, P_k$) as in the single mode case and convolve these elements together.
For instance, if we have $M$ identical but independent vibrational modes, the convolution over all $P_\pm$'s is straightforward to do and it only multiplies the Huang-Rhys factor $S$ by $M$, i.e., $S \rightarrow M S$.
Here, we assume this kind of a case: the Huang-Rhys factors can be large but we want to retain the tractability of the problem so we use only the minimal number of free parameters.

\section{Polaritonic spectrum}

Next, we discuss different features of the polaritonic spectrum that change due to vibrational dissipation.
The interest especially lies in the quantities that are inferred from experimental data: Rabi splittings, linewidths, and peak amplitudes.

\begin{figure}
    \centering
    \includegraphics{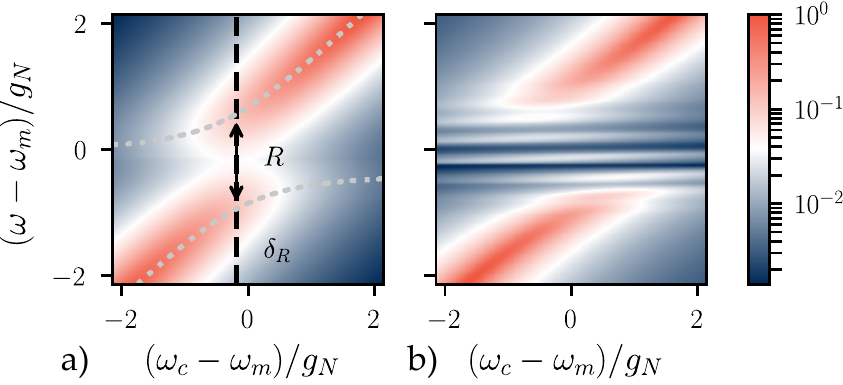}
    \caption{Polaritonic spectrum $S^T(\omega)$ for a) $Q = 0.9$ and $g_N = 7\omega_v$ and b) $Q = 5$ and $g_N = 3.5\omega_v$.
    In a), the gray dotted lines denote the position of the polariton peaks while the black dashed line represents the detuning $\delta_R$ on which the Rabi splitting $R$ is determined. 
    We set $S = 1, g_N/\kappa_c = g_N/k_B T = 3.5$, and the dissipation rate of the electronic transition  $\tilde{\kappa}/g_N = 0.01$ in both figures.
    }
    \label{fig:Qcomp}
\end{figure}

Vibrations provide an asymmetry to the absorption due to the emission-absorption asymmetry which manifests itself as a renormalization of the resonance condition.
In the coupled oscillator model, there are three distinct features at resonance $\omega_c = \omega_m$: 1) the difference between upper and lower polariton frequencies is at minimum, 2) the linewidths are the same and equal to the mean of cavity and exciton linewidths, and 3) the intensities of the peaks are equal.
In the presence of vibrations these conditions bifurcate to different values of detuning.
The role of Brownian vibrational dissipation to this ''bifurcation'' has not been investigated to our knowledge.
This phenomenon is often framed conversely as the asymmetry of the upper and lower polariton peaks which has garnered both 
theoretical~\cite{neuman2018origin,delpino2018tensor,reitz2019langevin,herrera2017theory,Herrera2017PRA,Herrera2017PRL,zeb2017exact,Michetti2008PRB,kansanen19} and 
experimental~\cite{bellessa2004strong,Chovan2008PRB,hakala2009vacuum,baieva2012strong,baieva2017dynamics} interest.
The notion of asymmetry often follows in theory even after fixing the detuning to zero by setting $\omega_c = \omega_m$. However, this might not give the minimum polariton peak separation, i.e., the Rabi split. 
This is demonstrated in Fig.~\ref{fig:Qcomp}a where the polariton spectrum is plotted as a function of the detuning and the driving frequency.

\subsection{Rabi splitting}
 
Let us first discuss in detail the renormalization of the observed polariton frequencies $\omega_\pm$ and especially the Rabi splitting $R$.
Here, we define the Rabi splitting as $R = \min_{\omega_c - \omega_m}(\omega_+ - \omega_-)$, and the value $\omega_c - \omega_m$ at which this minimum is obtained we denote by $\delta_R$.
To avoid ambiguities, we only look at parameter regimes where the vibrational sub-peaks are suppressed.
For instance, in Fig.~\ref{fig:Qcomp}b it is unclear, how the Rabi splitting should be evaluated.
This rarely is a problem in experiments with organic molecules.

\begin{figure}
    \centering
    \includegraphics{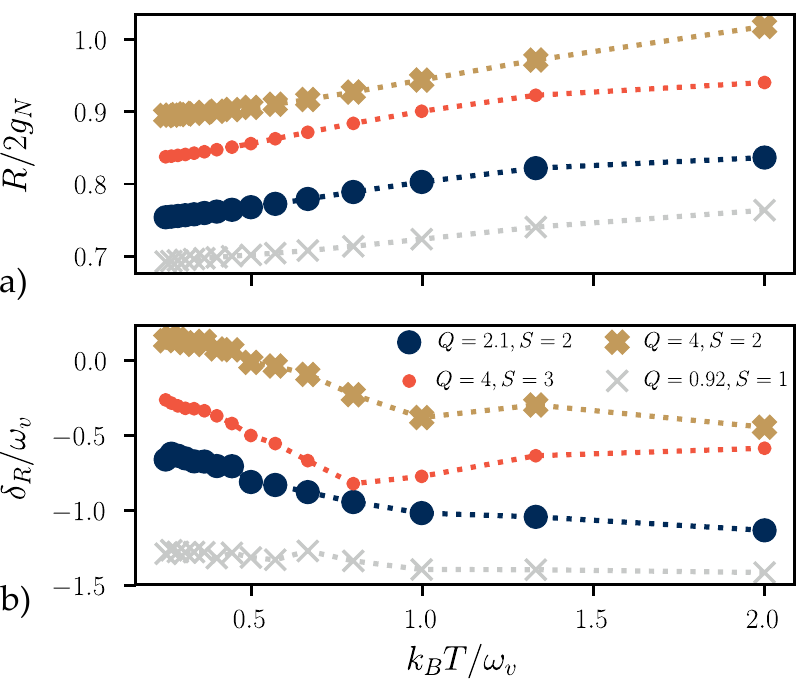}
    \caption{a) Rabi splitting $R$ and b) the corresponding detuning $\delta_{R}$ as a function of the temperature.
    The values of $Q$ and $S$ are chosen so that the absorption profile's linewidth at $k_B T = \omega_v$ are approximately equal for the circles and the crosses ($9.7\omega_v$ and $6.2\omega_v$ respectively).
    The temperatures are chosen in this figure so that the spacing between inverse temperatures is constant and each point corresponds to a different number of Matsubara modes.
    Here $g_N = 7\omega_v, \kappa_c = 2\omega_v, \tilde{\kappa} = 0.01\omega_v$.}
    \label{fig:rabisplits}
\end{figure}

We show in Fig.~\ref{fig:rabisplits}a how such an "experimentally inferred" Rabi splitting depends on the temperature $T$, vibration quality factor $Q$, and Huang-Rhys factor $S$.
As in the case of the coupled oscillator model, the Rabi splitting depends on the molecular linewidth. 
For example, $R$ increases with an increasing temperature which may be attributed to the increase in the overall linewidth.
However, the molecular lineshape plays an important role.
In Fig.~\ref{fig:rabisplits} we have chosen two pairs of values for $S$ and $Q$ so that there are two different effective linewidths indicated by the solid and dashed lines. 
These linewidths are determined as the full width at half maximum (FWHM) of the absorption profile $\Re(-A)$.
We find that the Rabi splitting is closer to the value expected from the coupled oscillator model ($2 g_N$ without any dissipation) with larger quality factors $Q$.

Contrary to the coupled oscillator model, the vibrational dissipation affects the relative position of the polariton frequencies to the uncoupled eigenfrequencies $\omega_m$ and $\omega_c$.
It is especially evident when the vibrations are of low quality.
This follows from our definition of $P(E)$: if the vibrations are very dissipative, $P(E)$ becomes centered around $E = 0$ because a vibronic peak at $E = \pm n \tilde{\omega}_v$ has the width of $n \gamma/2$.
Consequently, the absorption function $A(\omega - \omega_m + S \omega_v)$ in the response function and thus the absorption profile is peaked at $\omega = \omega_m - S \omega_v$.
This results to a shift of $S\omega_v$ of the polariton peaks which is most clearly seen as a shift in the local minimum of the polaritonic spectrum (see Fig.~\ref{fig:comp}).
If the vibrational quality factor is very high, $Q \gg 1$, then $\omega = \omega_m$ becomes again the absorption maximum.

An interesting consequence of the renormalization of polariton frequencies is that the detuning $\delta_R$ between bare cavity mode and the molecular exciton at which the Rabi splitting is determined changes.
This change is plotted in Fig.~\ref{fig:rabisplits}b as a function of temperature.
A part of the shift in the detuning by the vibrations is caused by the renormalization of the absorption frequency described in the previous paragraph.
However, even if this renormalization is taken into account, the effect still remains due to the change in
the molecular lineshape.
The frequency range of the change in detuning is of the order of one vibrational frequency.

\subsection{Linewidths}
Next, we consider the issue of polaritonic linewidth.
A natural point of comparison is the coupled oscillator model from which one infers the rule that polariton linewidth is the mean of cavity and molecular linewidth, although only at resonance $\omega_c = \omega_m$.
However, when vibrations are present, it is unclear what should be chosen as the molecular linewidth.
One option is the full width at half maximum (FWHM) of the molecular absorption spectrum.
Another option would be to deduce the value using a Lorentzian fit.
We choose the former method because of its simplicity.
Likewise, we use FWHMs as polariton linewidths.
To this end, we choose larger Rabi splittings so that the polariton peaks are clearly separated.

\begin{figure}
    \centering
    \includegraphics{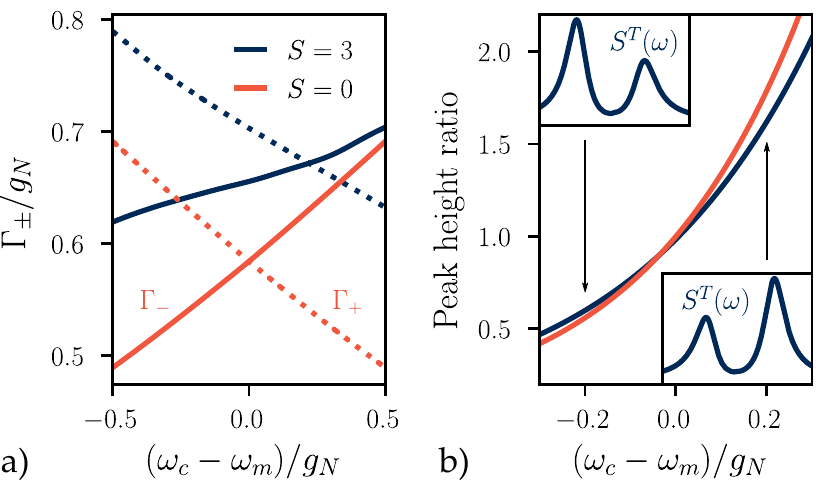}
    \caption{a) Estimated polariton linewidths $\Gamma_\pm$ (solid (dashed) line represents lower (upper) polariton) and b) the ratio of peak intensities of upper and lower polariton as a function of detuning.
    The insets show the polaritonic spectrum $S^T(\omega)$ corresponding to negative and positive detuning.
    Here, we have chosen the effective molecular linewidth (FWHM of absorption) as the dissipation rate $\kappa_m$ in the coupled oscillator model $S = 0$ (in orange). 
    The parameters are $Q = 4, g_N = 10\omega_v, \kappa_c = 0.2 g_N, \omega_v/k_B T = 1, \tilde{\kappa} = 0.06\omega_v$.}
    \label{fig:lw:detuning}
\end{figure}

In Fig.~\ref{fig:lw:detuning}a, we compare the polariton linewidths in the cases where the molecular broadening is Lorentzian (orange) or mostly caused by vibrations (blue).
We observe that the vibrations change the detuning $\delta_\Gamma$ at which the upper and lower polariton are of equal width $\Gamma_+ = \Gamma_-$.
Since at $\omega_c = \omega_m$ we have $\Gamma_+ > \Gamma_-$ as is well established, it is not surprising that $\delta_\Gamma > 0$.
The same effect may be seen in the ratio of upper/lower polariton peak intensities in Fig.~\ref{fig:lw:detuning}b, although to much lesser extent.
Together with the change in the detuning $\delta_R$ it appears that the ratio is below unity, i.e., the lower polariton peak is of higher intensity than the upper polariton at the detuning at which the Rabi splitting is determined.
This is in line with the previous experimental and theoretical analysis~\cite{neuman2018origin,delpino2018tensor,reitz2019langevin,herrera2017theory,Herrera2017PRA,Herrera2017PRL,zeb2017exact,Michetti2008PRB,kansanen19,bellessa2004strong,Chovan2008PRB,hakala2009vacuum,baieva2012strong,baieva2017dynamics}.
We also find that the polariton linewidths $\Gamma_\pm$ under vibrational dissipation are in general larger than those of the coupled oscillator model.
In our model, this is a consequence of the absorption function $A$ whose real part is even more tail-heavy than Lorentzian distributions.

\subsection{Limits of vibrational dissipation: overdamping and high quality vibrations}
So far, we have only considered the case in which the vibrations are underdamped and there exists a well defined vibrational frequency.
For completeness, we briefly comment the changes of the polaritonic spectrum induced by the overdamped vibrations.

\begin{figure}
    \centering
    \includegraphics{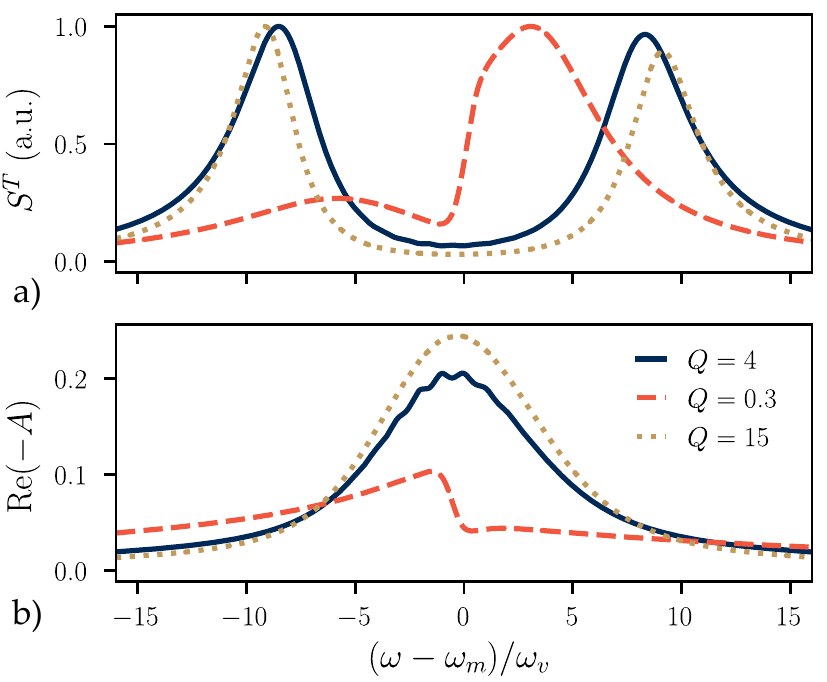}
    \caption{Three examples of different lineshapes that follow from the $P(E)$ theory to a) the polariton spectrum and b) absorption profile with the change of parameters. The blue line represents a Lorentzian-like case in which the vibrational dissipation dominates. The orange dashed line is in the overdamped limit while the brown dotted line is in the opposite limit of high quality vibrations.
    The parameters are chosen so that FWHMs are the same for each case, approximately $9.1\omega_v$.
    The vibrational parameters are $S = \qty{4,1,0.51}, \tilde{\kappa}/\omega_v = \qty{0.01,0.01,1.6}, \omega_v/k_B T = \qty{1, 0.56, 1}$ for $Q = \qty{4,0.3,15}$ respectively. The polariton spectrum is calculated with $g_N = 9\omega_v$ and $\kappa_c/g_N \approx 0.11$ and the detuning $\omega_c = \omega_m$.} 
    \label{fig:lineshapes}
\end{figure}

The effect of overdamping is that the absorption function becomes extremely tail-heavy as all the possible vibrational transitions are fully smeared by the dissipation.
Consequently, the emission-absorption asymmetry causes a very notable asymmetry to the absorption which is visible in the orange dashed line of Fig.~\ref{fig:lineshapes}.
Compared to the blue curve which is closer to a Lorentzian profile, the renormalization of the Rabi splitting becomes very large.
This is similar to what is observed in Fig.~\ref{fig:rabisplits} for low $Q$ values.
The change in the polariton lineshape is also notable but difficult to quantify.

We have considered here mainly the cases in which the vibrational dissipation is in fact the main source of dissipation.
However, it is possible that the vibrations are of very high quality but there is a large dissipation rate of the electronic transition.
In our model with only a single vibrational mode, this appears to be only way which leads to a Gaussian-like absorption profile.
It is the opposite limit to the overdamped vibrations as the tails of absorption profile become less pronounced than in Lorentzian profiles.
The effect on the results presented in Figs.~\ref{fig:rabisplits} and~\ref{fig:lw:detuning} is that the values change while the trends remain the same.
These changes may be inferred by comparing the brown dotted line in Fig.~\ref{fig:lineshapes} to the blue line that represents a Lorentzian-like case: 
the Rabi splitting are larger while the polariton linewidths smaller than in the corresponding Lorentzian case.

\section{Conclusions}
In summary, we have presented an analytical model on the effect of Brownian vibrational dissipation to the polariton spectrum in a cavity measurement.
Even though our approach is mathematically involved, it can provide richer physics than the coupled oscillator model.  
For most organic molecules, it should work as a much better approximation.
The associated cost is that there are a few parameters more regarding vibrations whose determination from experimental data requires more work.
However, our work provides some general checks to which the experimental data may be compared, such as the different detunings which give the Rabi splitting, equal polariton linewidths, and equal peak intensities. 

\begin{acknowledgments}
KSUK acknowledges the financial support of the Magnus Ehrnrooth foundation. This work was supported by the Academy of Finland projects HYNEQ, number 317118, and ManipuLight, number 323995. We thank Gerrit Groenhof for useful discussions.
\end{acknowledgments}

\section*{Data availability}
The data that support the findings of this study are available from the corresponding author
upon reasonable request.

\appendix

\section{Fourier transform of $\exp[\tilde{J}(t)]$}
\label{app:FT}

Formally, all the Fourier transforms in this work are of the form
\begin{align}
    \frac{1}{2\pi}\int \dd{t} e^{i E t} \exp[\qty(a + i b \sgn{t})\qty(e^{ i\omega_0 t - \Gamma \abs{t}} - 1)],
    \label{eq:app:start}
\end{align}
where $a, b, \omega_0, \Gamma \in \mathbb{R}$ and $\Gamma > 0$.
We use the convolution theorem to separate the Fourier transform into two parts before expanding the exponential into a Taylor series.
Thus, we have 
\begin{align}
    \frac{1}{2\pi}\int \dd{t} &e^{i E t} e^{- a - i b \sgn{t}} \notag\\
    &= e^{-a} \sum_{n=0}^\infty \frac{(- i b)^n}{n!} \frac{1}{2\pi}\int \dd{t} (\sgn{t})^n e^{i E t} \notag\\
    &= e^{-a}\qty[\delta(E) \cos{b} + \frac{\sin{b}}{\pi E}],
\end{align}
which holds true when treated as the Cauchy principal value integral.
At this point we can recognize that formally $\delta(E) = f_L(E;0,0)$ and $1/(\pi E) = g_L(E;0,0)$.
The second part of the transform is
\begin{align}
    \frac{1}{2\pi}\int \dd{t} e^{i E t} \exp[\qty(a + i b \sgn{t}) e^{ i\omega_0 t - \Gamma \abs{t}}] \notag\\
    = \sum_{n=0}^\infty \frac{1}{n!} \sum_{k=0}^n \mqty(n \\ k) I_k^n,
    \label{eq:app:premid}
\end{align}
where, with the help of the binomial theorem,
\begin{align}
    I^n_k = \frac{a^{n-k}}{2\pi}\int \dd{t}  (ib \sgn{t})^k \exp[i (E + n\omega_0) t - n\Gamma \abs{t}].
\end{align}
Again, the expressions with even $k$ differ from those of odd $k$.
By direct evaluation, one can confirm that
\begin{align}
    \frac{1}{2\pi}\int \dd{t} (\sgn{t})^k \exp[i (E + n\omega_0) t - n\Gamma \abs{t}] \notag\\
    = \begin{cases}
        f_L(E; - n \omega_0, n\Gamma) & \text{$k$ even,}\\
        i g_L(E; - n \omega_0, n\Gamma) & \text{$k$ odd.}
    \end{cases}
\end{align}
This result in conjunction with the convolution theorem leads directly to the convolution rules of $f_L$ and $g_L$ in Eq.~\eqref{eq:convrules}.
Furthermore, it can be used in $I^n_k$ which allows us to write the expression~\eqref{eq:app:premid} as
\begin{align}
     \sum_{n=0}^\infty \frac{1}{n!} \qty(f_L \sum_{\text{$k$ even}}^n + g_L\sum_{\text{$k$ odd}}^n) \mqty(n \\ k) a^{n-k} (i b)^k.
    \label{eq:app:midpoint}
\end{align}
We omit the arguments of $f_L$ and $g_L$ for brevity.
The even and odd binomial sums may be simplified by the relation
\begin{align}
    \frac{1}{2}[(a + ib)^n \pm (a - ib)^n] = 
    \sum_{\substack{\text{$k$ even} \\ \text{$k$ odd}}}^n \mqty(n \\ k) a^{n-k} (i b)^k,
\end{align}
which follows from the binomial theorem.
We denote now $z = a + i b = \abs{z} e^{i\theta}$ which simplifies the expression~\eqref{eq:app:midpoint} to
\begin{align}
    \sum_{n=0}^\infty \frac{\abs{z}^n}{n!} \qty[\cos(n\theta)f_L + \sin(n\theta)g_L].
\end{align}
The last step of convolving the results together is then straightforward with the convolution rules and by using the sum rules of trigonometric functions.
Thus, the Fourier transform~\eqref{eq:app:start} is
\begin{align}
    e^{-a}\sum_{n=0}^\infty \frac{\abs{z}^n}{n!} [&\cos(b - n\theta)f_L(E; - n \omega_0, n\Gamma) \notag\\
    &+ \sin(b - n\theta)g_L(E; - n \omega_0, n\Gamma)].
\end{align}
We find Eqs.~\eqref{eq:Pplusminus} and~\eqref{eq:Pmatsubara} of the main text by applying this result to $\tilde{J}_\pm$ and $\tilde{J}_k$.

\section*{References}
\bibliography{mainrefs.bib}

\end{document}